\documentclass{article}

\usepackage{PRIMEarxiv}

\usepackage[utf8]{inputenc} 
\usepackage[T1]{fontenc}    
\usepackage[authoryear]{natbib}  
\usepackage{hyperref}       
\usepackage{url}            
\usepackage{booktabs}       
\usepackage{amsfonts}       
\usepackage{nicefrac}       
\usepackage{microtype}      
\usepackage{fancyhdr}       
\usepackage{graphicx}       
\usepackage{amsmath,amssymb,amsthm}
\usepackage{subcaption,pgfplots,float}
\usepackage{xcolor}
\usepackage{lipsum}

\newtheorem{theorem}{Theorem}

\title{Optimal Experimental Design for Microplastics Sampling Experiments
\thanks{\textit{Citation:} M.A. Aquino-L\'opez, A.C. Ruiz-Fern\'andez, J.A. Sanchez-Cabeza, J.A. Christen. Optimal Experimental Design for Microplastics Sampling Experiments. 2025.}
}

\author{
Marco A. Aquino-L\'opez$^{\S 1}$~Ana Carolina Ruiz-Fern\'andez$^2$~Joan-Albert Sanchez-Cabeza$^2$~J. Andr\'es Christen$^1$ \\
$^\S$\texttt{aquino@cimat.mx} \\
\normalsize{$^1$ Centro de Investigaci\'on en Matem\'aticas (CIMAT-SECIHTI), Guanajuato, Mexico} \\
\normalsize{$^2$ Instituto de Ciencias del Mar y Limnolog\'ia, UNAM, Mazatl\'an, Mexico} 
}

\begin{document}
\maketitle

\begin{abstract}
Microplastics contamination is one of the most rapidly growing research topics. However, monitoring microplastics contamination in the environment presents both logistical and statistical challenges, particularly when constrained resources limit the scale of sampling and laboratory analysis. In this paper, we propose a Bayesian framework for the optimal experimental design of microplastic sampling campaigns. Our approach integrates prior knowledge and uncertainty quantification to guide decisions on how many spatial Centrosamples to collect and how many particles to analyze for polymer composition. By modeling particle counts as a Poisson distribution and polymer types as a Multinomial distribution, we developed a conjugate Bayesian model that enables efficient posterior inference. We introduce variance-based loss functions to evaluate expected information gain for both abundance and composition, and we formulate a constrained optimization problem that incorporates realistic cost structures. Our results provide principled and interpretable recommendations for allocating limited resources across the sampling and analysis phases. Through simulated scenarios and real-world-inspired examples, we demonstrate how the proposed methodology adapts to prior assumptions and cost variations, ensuring robustness and flexibility. This work contributes to the broader field of Bayesian experimental design by offering a concrete, application-driven case study that underscores the value of formal design strategies in environmental monitoring contexts.
\end{abstract}


\section{Introduction}\label{sec:intro}

Microplastics (MPs) contamination is a global issue of growing environmental concern. They are defined as plastic particles smaller than 5 mm and originate from various sources, including microbeads in personal care products, the breakdown of larger plastic debris, synthetic textiles, and tire abrasion, among many others \citep{andrady2011microplastics,gesamp2016sources}. Once MPs enter natural systems, they do not biodegrade but instead accumulate in various habitats, leading to long-term contamination \citep{rochman2013policy}. While their physical presence is well-documented, developing robust and cost-effective strategies to quantify their abundance and composition remains a key challenge, one that is inherently statistical in nature.

MPs pose significant threats to marine life, human health, and ecosystems. Aquatic organisms, ranging from plankton to large fish and whales, often mistake MPs for food, resulting in ingestion and bioaccumulation \citep{wright2017plastic}. These particles can transport harmful pollutants (i.e., persistent organic pollutants, or POPs, and heavy metals), exacerbating the dangers they present to wildlife and, potentially, humans through the food chain \citep{rochman2013policy}. 
These particles can transport harmful pollutants, such as persistent organic pollutants (POPs) and heavy metals, exacerbating the dangers they pose to wildlife and, potentially, humans through the food chain \citep{leslie2022discovery}. Nevertheless, risk assessment efforts—and the policy responses they inform, depend critically on the reliability of statistical inference drawn from often limited and resource-constrained sampling campaigns.

Given the immense global production and consumption of plastics, MPs are now ubiquitous in the environment, particularly in marine ecosystems \citep{andrady2011microplastics}. Coastal environments are hotspots for plastic debris, with beaches acting as sinks for MPs contamination carried by ocean currents and waves \citep{gesamp2016sources} and other sources.
The study of MPs contamination on beaches is important, as it provides a snapshot of broader environmental trends and enables targeted monitoring and remediation efforts. Understanding the concentration, distribution, and polymer composition of MPs in beach sand is essential to assess contamination levels, identify MPs sources, and evaluate potential impacts on local ecosystems and communities. These goals, however, require careful statistical planning to ensure that sampling schemes are both efficient and capable of maximizing information gain, which can be represented, for example, as variance ratio reduction in a Bayesian framework \citep{bernardo1994, Ryan_2015}. Choices regarding the number of samples to collect and how to allocate laboratory resources directly influence the quality and interpretability of the resulting data.

Monitoring MPs faces several challenges. Analysis is expensive and time-consuming, requiring the counting of suspected MPs particles and, in turn, the determination of polymer composition to confirm their plastic nature and likely identify contamination sources \citep{hidalgo2012microplastics}. 
Globally harmonized protocols for sample collection and laboratory analysis are essential to ensure consistent, reliable and intercomparable results.
Key methodological limitations include properly reporting uncertainties in abundance, determining the sample size for analysis, and establishing the number of particles for polymer identification, all while maintaining a cost-benefit balance \citep{hermsen2018quality}.
These limitations highlight the need for a formal statistical approach that explicitly accounts for uncertainty, resource constraints, and the trade-offs between sampling breadth and analytical depth.

In this study, we present a methodology based on a Bayesian approach to address MPs abundance, polymer categorization estimates, and their uncertainties.
More importantly, we proposed a framework to establish a cost-effective sampling strategy and polymer categorization based on a formal decision theoretic sequential analysis.
Our methodology is grounded in the optimization of loss functions, allowing for principled design of sampling campaigns that balance statistical precision and operational cost \citep{ryan2016}. 

We expect that this contribution will enhance the understanding of MPs contamination by providing a protocol that designs an organization (e.g. \url{https://remarco.org/}, see below) to create a framework for planning cost-effective sampling campaigns, standardizing criteria and reporting protocols, and incorporating correct uncertainty quantification of estimates. While the Bayesian inference procedure we propose for estimating abundance and polymer composition is conceptually straightforward (a conjugate analysis), the primary aim of this article is to provide guidelines for an optimal experimental design \citep{muller2005, DeGroot1970, chaloner1995}.  A novel formal sequential analysis will be followed, to plan first for sample collection and, in turn, for polymer classification sub-sampling.

The paper is organized as follows.
In Section~\ref{sec:current_prac} we present current practices in sampling and data reporting MPs counting and categorization, specifying the characteristics and basic assumptions of the sequential experimental design problem to be addressed in the paper.
Section~\ref{sec:model} presents the full probabilistic model and its conjugate structure, which enables efficient posterior computation and inference. In Section~\ref{sec:costs}, we introduce loss functions that quantify the expected reduction in posterior uncertainty for both abundance and composition, and combine them with a realistic cost model to formulate a constrained optimization problem, to solve the sequential experimental design problem. Section~\ref{sec:illustration} illustrates the practical behavior of our design strategy through simulations under varying scenarios, and Section~\ref{sec:examples} provides an application using empirical data from a beach monitoring campaign. Finally, Section~\ref{sec:discussion} draws some conclusions of the paper and outlines future work.

\section{Current practices in MPs counting and polymer categorization}
\label{sec:current_prac}

As an example, we focus on the harmonized method developed by the Research Network of Marine-Coastal Stressors in Latin America and the Caribbean \citep[REMARCO;][]{REMARCO2024} with support from the International Atomic Energy Agency (IAEA). This method aims to improve data comparability and reliability across MPs monitoring of sand beaches in 18 countries. The REMARCO method involves collecting five sand samples per beach, using 50 × 50 cm quadrats, each subdivided into four 25 × 25 cm quadrants. These quadrants are positioned at 20 m intervals, along a 100 m transect established parallel to the coastline at the most recent high tide line.

Sand samples, consisting of the top 1 cm of  sediment, are collected with a stainless-steel spoon or shovel from a randomly selected 25 × 25 cm quadrant, within each quadrat. Each sample is stored in a glass or metal container and maintained at room temperature until analysis.
In the laboratory, MPs are extracted through density separation, using a saturated NaCl solution (1.2~g~cm$^{-3}$) and recovered by filtration in 250 $\mu$m metallic filters.
Filters are examined under a stereomicroscope (40X magnification) to count and classify the suspected MPs
\citep[see][for details]{REMARCO2024}.
Once the suspected MPs have been counted,
for confirmation and polymer categorization, a minimum of 100 particles per replicate (or all particles if fewer than 100) are analyzed using ATR-FTIR or Raman spectroscopy. The experimental design strategy of sampling 5 quadrants and sub-sampling ``a minimum'' of 100 MPs for FTIR or Raman categorization are given in REMARCO \citep{REMARCO2024} as guidelines, without any formal statistical analysis. 

Sand samples, consisting of the top 1~cm of sediment, are collected with a stainless-steel spoon or shovel from a randomly selected 25~×~25~cm quadrant, within each quadrat. Each sample is stored in a glass or metal container and maintained at room temperature until analysis. In the laboratory, MPs are extracted through density separation, using a saturated NaCl solution (1.2~g~cm$^{-3}$) and recovered by filtration in 250-$\mu$m metallic filters. Filters are examined under a stereomicroscope (40X magnification) to count and classify the suspected MPs \citep[see][for details]{REMARCO2024}. Once the suspected MPs have been counted, for confirmation and polymer categorization, a minimum of 100 particles per replicate (or all particles if fewer than 100) are analyzed using Fourier Transform Infrared (FTIR) or Raman spectroscopy \citep{Hossain2020}. Despite the REMARCO's method representing a significant improvement harmonization effort, that is currently being transferred to other regions of the world through IAEA technical cooperation projects, the experimental design strategy of sampling five quadrants and sub-sampling ``a minimum'' of 100 MPs for FTIR or Raman categorization \citep{REMARCO2024} have been developed without formal statistical foundation.

MPs abundance is calculated with
\begin{equation}
\text{Abundance of MP (MP m}^{-2}) = \frac{\#\text{total MPs in the filter}}{\text{Sample collection area}}.\label{eq:esti1}
\end{equation}
Equation~\ref{eq:esti1} shows the estimator currently used for MPs abundance. Then, using the mean value of the MPs counts from the five quadrants (each of them considered to be a replicate) and a simple standard deviation, a ``confidence interval'' is calculated. However, this procedure can be improved since confidence intervals often include negative values, which are physically meaningless. Moreover, polymer classification proportions are reported directly from the observed counts \citep[e.g.,][]{GarciaVarens2025}.

These inconsistencies highlight the need for a formal inferential framework that yields coherent estimates and properly accounts for uncertainty. It is worth noting that such a statistical framework is not inherently complex, but it does require a formal statistical approach.

As FTIR or Raman spectroscopy for polymer identification is costly and time-consuming, our approach aims to determine the proportion of suspected MPs that need to be characterized to achieve a meaningful and cost-effective inference. Common polymers of interest (mass-produced plastics, \citealt{Geyer2017}) are shown in Table~\ref{tab:polymer_types}.

\begin{table}[h!]
\centering
\caption{Ten Polymer Types of interest Found in Microplastic Samples}
\label{tab:polymer_types}
\begin{tabular}{l l | l l}
\toprule
\textbf{Abbr.} & \textbf{Polymer Name}
& \textbf{Abbr.} & \textbf{Polymer Name}\\
\midrule
PE  & Polyethylene  & PVC & Polyvinyl Chloride \\
PP  & Polypropylene & PU  & Polyurethane\\
PET & Polyethylene Terephthalate & AC  & Acrylic\\
PS  & Polystyrene & PES & Polyester\\
PA  & Polyamide & NPP & Non-Plastic Polymer \\
\bottomrule
\end{tabular}
\end{table}
To rigorously support and optimize this monitoring, we introduce a Bayesian statistical framework that characterizes the process of sampling MPs on beaches through two steps: estimating the total abundance and classifying particles into polymer categories. Specifically, we model the count of suspected MPs in each sampled area as a Poisson process governed by an underlying Gamma-distributed abundance rate, and the categorization of these particles as a Multinomial process, with a Dirichlet prior, reflecting polymer composition. This modeling choice reflects both the hierarchical structure of the data and the inferential goals of estimating the quantities of interest under uncertainty.

For sampling MPs on a beach, we build on existing field protocols, namely randomization of quadrants \citep[REMARCO;][]{REMARCO2024}. Although our study focuses on beach environments, where sampling protocols already exist, the proposed methodology is general and may be applied to other settings, such as marine sediments, lakebeds, peatlands, or ice cores with minor modifications. The statistical modeling and the specific experimental design will be explained in the next section.


\section{Model formulation, inference, design problem and expected variance reduction}
\label{sec:model}
This chapter introduces the statistical framework used to model the two key components of our analysis. The first step focuses on estimating the abundance of MPs based on observed particle counts. The second step involves categorizing the identified particles into polymers of interest. We formulate a Bayesian model that links both stages, providing a coherent structure for inference that is the basis of our experimental design.

We focus our efforts on the specific case of MPs sampling on beaches, although this analysis can be easily modified for other cases. For a particular site, let $N_j$ denote the total number of suspected microplastics in sample $j$ in a predetermined area of size $A$ (in m$^2$). We model $N_j$ as a Poisson random variable (r.v.), namely
\begin{align*} 
N_j \mid \Lambda = \lambda ~\sim \text{Poi}(A\lambda), 
\end{align*}
where $j = 1, 2, \dots, m$. The key design variable is $m$, while $\Lambda$ is the parameter of interest for the first step of the process. 
We assume that each $N_j$ is conditionally independent, representing an observation from the same underlying population. Each count is conditional on the sampled area $A$ and the suspected microplastics density $\Lambda$, expressed in units of suspected MPs per unit area (MP~m$^{-2}$). 
This assumption aligns with the standard sampling strategy, in which the site is divided into 1~m$^2$ \emph{quadrats}, each subdivided into four equal 0.25~m~$\times$~0.25~m \emph{quadrants}. One quadrant is randomly selected from each quadrat, resulting in $m$ sampled quadrants, each with a predetermined area $A$ (current standards indicate $A = 0.25^2$~m$^2$). Note that each quadrat contributes exactly one sampled quadrant; therefore, from this point onward, we will refer simply to \emph{quadrants}.

Up to this point, the observed particles are only suspected to be microplastics. To determine whether a given particle is indeed a MP, a categorization analysis is required. For this reason, the next step in our inferential framework is to classify the suspected particles. We assume the existence of $k$ categories: $k-1$ corresponding to distinct types of plastic polymers of interest, and one representing ``non-plastic'' materials (e.g., natural polymers and other non-plastic substances). In typical applications, it is possible to assume $k = 10$; see Table~\ref{tab:polymer_types}. From this point onward, we refer to this step as polymer categorization.

Polymer categorization always occurs after the MPs counting process; therefore, the values $N_j$ are already observed at this point. This implies that, conditional on $N_j = n_j$, the particle classification follows a Multinomial model, namely
\begin{align*} 
\mathbf{S}_j \mid \mathbf{P} = \mathbf{p}, N_j = n_j \sim \text{Mult}( n_j, \mathbf{p}), 
\end{align*}
where $\mathbf{S}_j = (S_{1j}, S_{2j}, \dots, S_{kj})$ are the $k$ categories counts and $\mathbf{P} = (P_1, P_2, \dots, P_k)$ is the vector of categories proportions; certainly $N_j = \sum_{i=1}^k S_{ij}$ and $\sum_{i=1}^k  P_i = 1$.
This ensures that the model is coherent with the assumption that MPs from one category should not provide information about any other category.

With these assumptions, we may now establish a joint model for all data. Let $\mathbf{N} = (N_1, \ldots , N_m)$, \begin{align}
\label{eqn:joint_model}
f_{\mathbf{S}_1, \ldots, \mathbf{S}_k, \mathbf{N} \mid \Lambda, \mathbf{P}}( \mathbf{s}_1, \ldots , \mathbf{s}_k, \mathbf{n} \mid \lambda, \mathbf{p}, m )  &=
\prod_{j=1}^{m}  f_{\mathbf{S}_j \mid N_j, \mathbf{P}}( \mathbf{s}_j \mid n_j, \mathbf{p}) f_{ N_j \mid \Lambda}( n_j \mid \lambda, A) \nonumber \\
&= \prod_{j=1}^{m} \text{Multi}( \mathbf{s}_j  \mid \mathbf{p},n_j) \text{Poi}(n_j \mid \lambda  A ) . 
\end{align}

Now, we specify prior distributions for the model parameters, namely $\Lambda$ and $\mathbf{P}$. The abundance rate parameter $\Lambda$ is a strictly positive quantity, and a Gamma distribution offers a flexible and interpretable prior choice with domain on $\mathbb{R}^+$, which make it suable for modeling this parameter and indeed, this is also chosen for mathematical convenience since it is a conjugate prior for the assumed Poisson distribution of the $N_j$s, that is, 
\begin{align*} 
\Lambda \sim \text{Gamma}(\alpha, \beta),
\end{align*}
where $\alpha$ and $\beta$ are the prior shape and \textit{rate} parameters.  

We know that vector $\mathbf{P}$ belongs in the standard simplex $k-1$. For this reason, we choose a Dirichlet distribution, namely
\begin{align*} 
\mathbf{P} \sim \text{Dir}_k(\gamma),
\end{align*}
where $\gamma$ is a vector with $k$ elements and reflects our prior knowladge about the proportions of categories of suspected MP. 
In this parametrization we assume $\sum_{i=1}^k \gamma_k = 1$.

With these prior specifications it is clear that the posterior may be written as,
\begin{align*}
& f_{\Lambda, \mathbf{P} \mid \mathbf{S}_1, \ldots , \mathbf{S}_k, \mathbf{N}}(\lambda, \mathbf{p} \mid \mathbf{s}_1, \ldots , \mathbf{s}_k, \mathbf{n}, A )  \\
& \propto f_{\mathbf{P}}(\mathbf{p)}
\prod_{j=1}^{m}  f_{\mathbf{S}_j \mid N_j, \mathbf{P}}( \mathbf{s}_j \mid n_j, \mathbf{p})
\cdot f_{\Lambda}(\lambda) \prod_{j=1}^{m} f_{ N_j \mid \Lambda}( n_j \mid \lambda, A) \\
&\propto \text{Dir}_k(\mathbf{p} | \gamma ) \prod_{j=1}^{m} \text{Multi}( \mathbf{c}_j  \mid \mathbf{p},n_j) \cdot \text{Gamma}(\lambda | \alpha, \beta) \prod_{j=1}^{m} \text{Poi}(n_j \mid \lambda  A ) . 
\end{align*}

At this point, we can exploit conjugacy to simplify the posterior distribution. The second term corresponds to the posterior for $\Lambda$, which follows a $\text{Gamma}(\alpha + n, \beta + mA)$ distribution, where $n = \sum_{j=1}^m n_j$ is the total number of individual microplastics that were categorized. Similarly, the first term yields a Dirichlet distribution for $\mathbf{P} \sim \text{Dir}_k\left(\gamma_1 + s_1, \dots, \gamma_k + s_k\right)$,
where $s_i = \sum_{j=1}^m s_{ij}$ represents the total count of microplastics classified as type $i$ across all samples.  
Let $\mathbf{s} = (s_1, \dots, s_k)$; the full posterior distribution may then be written as
\begin{align} 
f_{\Lambda, \mathbf{P} \mid \mathbf{S}, N}(\lambda, \mathbf{p} \mid \mathbf{s}, n, m A )
&= f_{\mathbf{P} \mid \mathbf{S}, N}(\mathbf{p} \mid \mathbf{s}, n)~
f_{\Lambda \mid N}(\lambda \mid n, m A ) \nonumber \\ 
&= \text{Dir}_k\left(\mathbf{p} \mid \gamma + \mathbf{s}, n\right) ~\cdot~ \text{Gamma}\left(\lambda \mid \alpha + n, \beta + mA\right) 
\label{eq:fulL_posterior}
\end{align}
with the previously defined notation.
Note that all normalization constants are explicitly known, justifying the use of the equality sign in the expression for the posterior. Furthermore, the joint posterior no longer depends on the individual values of $N_j$, but only on the total number of suspected microplastics found across all samples, $N = \sum_{j=1}^m N_j$, and the total sampled area, $mA$.

This result provides posterior inference on the parameters $\Lambda$ and $\mathbf{P}$ and is the basis for our experimental design. The number of quadrants, $m$, is to be decided before sampling, and a proportion of suspected MPs should be selected for polymer categorization. Let $q \in [0, 1]$ represent this proportion. We define $\bar{n}(q) = \lfloor nq \rfloor$ as the actual number of individual particles sent for polymer categorization. The task, therefore, is to establish a loss function to determine the optimal $m$ and, conditional on the total number of MPs, $n$, to decide the proportion $q$ for polymer categorization. That is, a sequential experimental design. For $\Lambda$ and $\mathbf{P}$, the loss (or information gain from sampling) is defined as the ratio between the prior and posterior variance (see Sections~\ref{sec:lossfun_1} and~\ref{sec:lossfun_2}). Cost considerations and the overall sequential optimization design are explained in Section~\ref{sec:costs}.

\subsection{Expected variance reduction in MPs count}\label{sec:lossfun_1}

In this section, we construct the loss function used to optimize the number of samples $m$ in the experimental design for estimating the MPs abundance parameter $\Lambda$. Our objective is to quantify the expected reduction in posterior variance relative to the prior, as a function of the sampling effort determined by $m$. This leads to the following variance-based loss function:
\begin{equation}
L_1( m\mid N,\alpha,\beta,A) = \frac{\text{Var}(\Lambda \mid N, \alpha, \beta, m, A)}{\text{Var}(\Lambda \mid \alpha, \beta)}. \label{eq:loss_1}
\end{equation}
This loss function can be formally justified under a quadratic loss for estimating $\Lambda$, as discussed in \citet{bernardo1994}. We do not adopt this loss function to target a specific decision or hypothesis test (e.g., distinguishing between human-derived and natural sources of contamination). Instead, our loss functions should be seen as neutral or generic tools for further inference, aimed at maximizing information gain through variance reduction in the posterior distributions of the parameters. Once inference is complete, the resulting posterior distributions can be used for various downstream analyses, including formal hypothesis testing or threshold-based decision-making.

Note that the loss function in Equation~\eqref{eq:loss_1} is conditional on the (as yet unobserved) sample counts $\mathbf{N}$. To evaluate the design prior to data collection, we therefore consider the \textit{prior expected} reduction in variance. The following result, proved in Appendix~\ref{sec:appendix}, provides a closed-form expression for this expectation.

\begin{theorem}
\label{th:ExpVarLa}
The prior expected reduction in variance of $\Lambda$ is
\begin{align}
	L_1^*( m | \alpha, \beta, A ) = \frac{1}{1+\frac{\nu_0}{\lambda_0} m A}
\end{align}    
where $\lambda_0 = \frac{\alpha}{\beta}$ and $\nu_0 = \frac{\alpha}{\beta^2}$ are the prior mean and variance of $\Lambda$, respectively.
\end{theorem}

\subsection{Expected variance reduction in classification of microplastics}\label{sec:lossfun_2}

Assume we randomly select a proportion $q \in (0,1)$ of the total observed particles to construct a sub-sampled count matrix $\mathbf{S}$. The model assumptions remain valid under this sub-sampling scheme when computing the posterior distribution of the class probabilities.

Let $\bar{s}_i = \sum_{j=1}^m \bar{s}_{ij}$ denote the total number of MP of type $i$ in the sub-sample, and define the sub-sample count vector as $\bar{\mathbf{s}}(q) = (\bar{s}_1, \ldots, \bar{s}_k)$. The total number of microplastics selected for categorization is denoted by $\bar{n}(q) = \left\lfloor n q \right\rfloor$, where $n = \sum_{j=1}^m n_j$ is the total number of suspected microplastics. Then, certainly, the posterior distribution for $\mathbf{P}$ under the sub-sampled data remains Dirichlet with updated counts $\bar{\mathbf{s}}(q)$, preserving conjugacy, namely
\begin{align} 
f_{\mathbf{P} \mid \mathbf{S}, N}(\mathbf{p} \mid \bar{\mathbf{s}}(q), \bar{n}(q))
     &= \text{Dir}_k \left(\mathbf{p} \mid \gamma + \bar{\mathbf{s}}(q), \bar{n}(q) \right).\label{eq:p_posterior} 
\end{align}

We define the expected reduction in variance as  
\begin{align}
	 L_2(q \mid \bar{\mathbf{s}}, n, \boldsymbol{\gamma}) 
     &= \frac{\text{Tr}(\Sigma_1)}{ \text{Tr}(\Sigma_0) },
     \label{eq:L_star_p_nqc}
\end{align}
where $\Sigma_0$ and $\Sigma_1$ denote the prior and posterior covariance matrices of $\mathbf{P}$, respectively, using the subsample.
As for $L_1$, this loss can be formally justified under a squared error loss for a point estimator of $\mathbf{P}$ \citet{bernardo1994}.

The expected reduction in variance depends on the as-yet unobserved category counts $\mathbf{S}$. To assess the design before data collection, we require the \textit{prior expected} reduction in variance. The following theorem, which provides a closed-form expression, is proved in Appendix~\ref{sec:appendix}.
\begin{theorem}
\label{th:ExpVarP}
The prior expected reduction in variance for $\mathbf{P}$ is
\begin{align}
L_2^*( q \mid n, \mathbf{\gamma})
&= \frac{\gamma_0 + 1 - \frac{\bar{n}(q)}{\gamma_0 + \bar{n}(q)}}{\gamma_0 + 1 + \bar{n}(q)}
\end{align}
where $\gamma_0 = \sum_{i=1}^k \gamma_i$.
\end{theorem}

\section{Cost Function and Complete Sequential Design}\label{sec:costs}

Any research study that requires data collection is inevitably constrained by limited resources in terms of time, labor, and financial cost. The design of a sampling effort must carefully weigh the trade-off between the information gained and the resources required. While collecting the largest possible sample might be ideal in theory, practical constraints necessitate the explicit modeling of costs and the development of an optimized, resource-efficient sampling strategy. 

In our context, the gain in knowledge is quantified by the expected reduction in variance of the key parameters $\Lambda$ and $\mathbf{P}$. On the other hand, costs arise at multiple stages: during field sampling, pre-processing, and counting, as well as the more resource-intensive task of polymer categorization. In this section, we introduce a general structure for incorporating these cost components into the experimental design, enabling a cost-benefit analysis.

In the context of MPs field studies, the total cost of an experiment typically comprises multiple components, which can be broadly categorized into fixed costs, field-related variable costs, and laboratory processing costs. ``Fixed costs'' refer to baseline expenditures that are incurred regardless of the sampling effort while field-related variable costs, by contrast, are those directly tied to the sampling effort.

Laboratory costs represent a third major component and include both manual and instrument-based analysis. Personnel time spent under a microscope—used to count and visually identify suspected MPs—can be substantial, especially when large numbers of samples are involved. Finally, polymer analysis using spectroscopic methods, such as FTIR and Raman, incurs an additional cost per particle, reflecting both machine time and operator labor.

These distinctions are not only useful for understanding which costs arise at different stages of the process—sampling, counting, and polymer analysis—but also serve to help researchers identify which components are suitable for optimization. While some costs are unavoidable and fixed, others vary depending on design choices, such as the number of samples or the area covered. In our study, we represent the total cost using the following decomposition:
\begin{equation}
c_0 + m(c_1 + c_2 A) + c_3 n + c_4 \bar n(q),
\end{equation}
where each term corresponds to a specific aspect of the process. The constant $c_0$ captures fixed baseline costs, such as general setup and field logistics, which are incurred regardless of sampling intensity. The term $m(c_1+c_2A)$ reflects the cost of field sampling, combining a per-sample cost $c_1$ (e.g., labor, transportation) with a cost per unit area $c_2$ related to the effort of covering and processing the sampled sediment. The term $c_3~n$ represents laboratory effort associated with counting all suspected MPs under a microscope. Finally, $c_4~\bar n(q)$ accounts for the cost of polymer analysis, which is applied only to the sub-sample of particles selected for polymer categorization.

In many field-based studies, especially those involving spatially distributed sampling, the cost per sampled area ($c_2A$) tends to dominate the per-sample cost ($c_1$). When this is the case, and under the assumption that $c_0$ is a fixed sunk cost unrelated to the design, the total cost function can be simplified by excluding both $c_0$ and $c_1$. To further clarify the trade-off between spatial coverage  and analytical resolution, we factor out $c_2$ to express the total cost in the following form
\begin{equation}
\label{eq:cost_0}
C_0(A, m, q, n ) = c_2 \left[ m A +  n \left( \frac{c_3}{c_2} \right) + \left( \frac{c_4}{c_2} \right) \bar{n}(q) \right].
\end{equation}
This expression highlights how the total cost depends on three components: the sampling effort across space ($mA$), the effort required to count all observed MPs ($n$), and the cost of characterizing a selected subset of suspected MP particles ($\bar{n}(q)$) through FTIR or Raman analysis.

While we refer to $c_2$, $c_3$, and $c_4$ as costs, it is important to note that these terms do not solely represent monetary expenditures. Instead, they reflect a broader notion of resource use such as; time, labor, and logistical effort associated with each phase of the analysis. To make these trade-offs more interpretable and comparable across studies, we define the following dimensionless ratios
\begin{equation}
r_1 =  \frac{c_3}{c_2} ~~\text{and}~~ r_2 =  \frac{c_4}{c_2} .
\end{equation}
These ratios quantify the relative efforts related to field sampling. Specifically, $r_1$ represents the cost of counting one suspected MP particle relative to sampling one square meter of sand, while $r_2$ denotes the relative cost of categorizing a single suspected MP when compared to sampling one squared meter of sand. This normalization serves to homogenize the cost components by removing dependence on specific currencies, local economic conditions, or institutional contexts, thereby providing a standardized notion of relative effort across phases of the analysis.

Once fixed costs have been considered, the remaining resources must be distributed across the various stages of the study: sampling, particle counting, and polymer categorization. Let $b_t$ denote this remaining budget. To express the remaining cost in relative terms and facilitate comparison across designs, we normalize the total cost \eqref{eq:cost_0} with respect to $b_t$, yielding the dimensionless expression
\begin{equation}
C( m A, q, n ) = 
c \left[ m A + r_1 n + r_2  \bar{n}(q) \right] ,
\end{equation}
where $c=c_2/b_t$.
The constant $c$ is the proportion of the available budget required to sample one square meter of sand. This normalized formulation is independent of currency and institutional context, allowing for direct interpretation of the trade-offs between field and laboratory efforts. To remain feasible, any design must satisfy the constraint $C(mA,q,n)\leq1$, ensuring that the total cost does not exceed the available budget.

Given this parametrization, we emphasize that the goal of the design is not to minimize cost \textit{per se}, but rather to determine how to allocate the available resources efficiently. That is, we assume that the total available resources have already been established, and our objective is to guide how this budget should be distributed between sampling and polymer categorization.

From this perspective, the cost function is treated as a constraint rather than an objective to minimize. Therefore \textit{we assume that any remaining resources after the field sampling stage should be fully allocated to the categorization of microplastics, until the budget is exhausted}. This leads to the following constraint formulation
\begin{equation}
\label{eq:constraint}
C( mA, n, q) = 1, q < 1, ~~\text{and}~~ C(  mA, n, 1) \leq 1 .
\end{equation}
The first condition requires full budget utilization when only a proportion of particles is categorized ($q < 1$), while the second permits the possibility that all particles are categorized ($q = 1$) without necessarily exhausting the budget.

Condition~\eqref{eq:constraint} defines the feasible region of all admissible combinations of total sampled area $mA$ and the number of categorized particles $\bar n(q)$ under the fixed budget. 
Condition~\eqref{eq:constraint} implicitly defines $q$ as a function of the sampling effort $mA$ and the total number of observed particles $n$. Specifically, we have from \eqref{eq:constraint}
\begin{equation}
q( m A, n) = \max \left( 0, \min \left( 1, \frac{1/c - ( m A + n r_1)}{r_2 n} \right) \right).\label{eq:q}    
\end{equation}

This formulation defines the design decision as a constrained optimization problem.  As constructed, the experimental design is conducted in two sequential steps. First, a decision is made regarding the total area to be sampled ($m A$). Second,  \textit{conditional} on the total number of suspected MP $n$ found in the sample area, the categorization proportion $q$ is determined by using the budget constrain \eqref{eq:constraint}, that is, $q( m A, n)$.
It is then clear that the overall \textit{expected} cost is
\begin{equation}
\label{eq:total_L}
L^*( m \mid A ) =
\frac{1}{2} L_1^*( m \mid A, \alpha, \beta ) +
\frac{1}{2} E_{N | \alpha, \beta, m A}[ L_2^*( q \mid N , \mathbf{\gamma} )] .
\end{equation}
That is, the expected cost of analyzing the suspected MPs—corresponding to the second stage of the sampling campaign—is evaluated \textit{conditional} on having already selected the number of quadrats $m$ to be sampled. To compute this expectation, we use the predictive distribution of the total count $N$, obtained from the Poisson–Gamma model described in Section~\ref{sec:model}.
We multiply the sum by $\frac{1}{2}$ only to maintain $L^*$ also in $[0,1]$.

Then, combining \eqref{eq:total_L} with condition \eqref{eq:constraint} the optimization task is
\begin{equation}
\min_{m \in \mathcal{A}} L^*( m \mid A ) =
\min_{m \in \mathcal{A}} \left\{ \frac{1}{2} L_1^*( m, A \mid  \alpha, \beta ) + \frac{1}{2} E_{N | \alpha, \beta, m A}[ L_2^*( q \mid N , \mathbf{\gamma} )] \right\}  ,
\label{eq:opti_L}
\end{equation}
where $m \in \mathcal{A} = \{ 0,1, \ldots,  \lfloor (A c)^{-1} \rfloor \}$ ensures not to exceed the total budget.
This overall expected cost function may be easily justified in terms of a formal two stage Backward-Induction procedure \citep[][chap. 12]{DeGroot1970}. The corresponding minimisation in the second stage not explicitly done due to the constrain \eqref{eq:constraint}.

The expectation in Equation~\eqref{eq:opti_L} cannot be computed analytically. However, since the predictive distribution of $N \mid \alpha, \beta, mA$ is a Poisson-Gamma (see Appendix~\ref{sec:appendix}), it can be easily simulated, and with a large sample, a very good approximation of the expectation can be obtained. This needs to be done for each $m$. Typically, the maximum number of quadrants, $m$, is around 15 to 20, and the calculation of all values of $L^*(m \mid A)$ can be done in less than 1 minute in our programme.

It is important to note that, from the outset, we do not address the design of the randomized sampling area within each quadrat, denoted by $A$, which is assumed to be fixed and standardized throughout the study.

\section{Performance illustration of the sequential design}\label{sec:illustration}

In this section, we intend to illustrate the performance of the sequential design with two examples. In analyzing previous sampling campaigns on beaches near Mazatlán, Mexico \citep{Rios_Mendoza_2021}, we established a total budget, after removing fixed costs, equivalent to 12 quadrants that could be sampled on the beach. In any case, this figure should be regarded solely as an illustrative value for the examples in this section.

Each quadrant samples a sand area of $A = 0.25^2$ m$^2$, thus $c = ( 12 A )^{-1}$.  In rounded numbers, the corresponding proportion of counting one MP in the stereomicroscope was established as $r_1 = 5 \times 10^{-5}$ and the corresponding proportion of categorizing a MP through polymer analysis as $r_2 = 3 \times 10^{-3}$, both compared to sampling the 12 quadrants of beach (the total budget).  We proceed with these values and compare the resulting sampling designs for two different prior distributions for the MP abundance $\Lambda$.  

\begin{figure}[h!]
\begin{center}
\begin{tabular}{cc}
\includegraphics[scale=0.4]{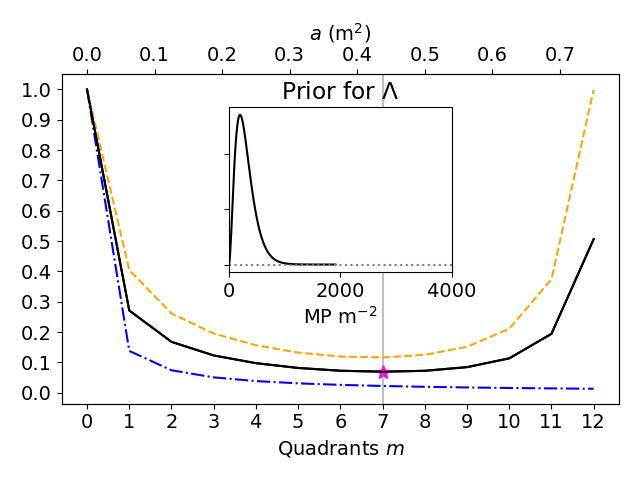} &
\includegraphics[scale=0.4]{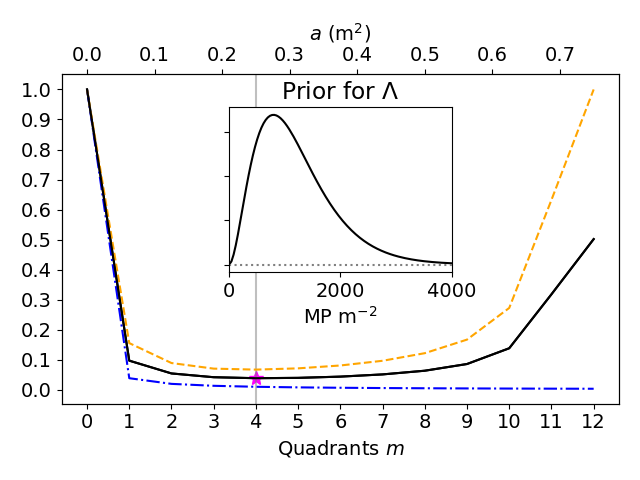} \\
\includegraphics[scale=0.4]{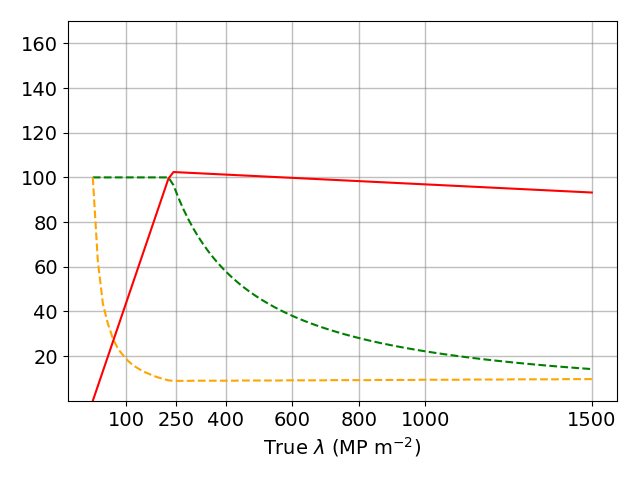}  &
\includegraphics[scale=0.4]{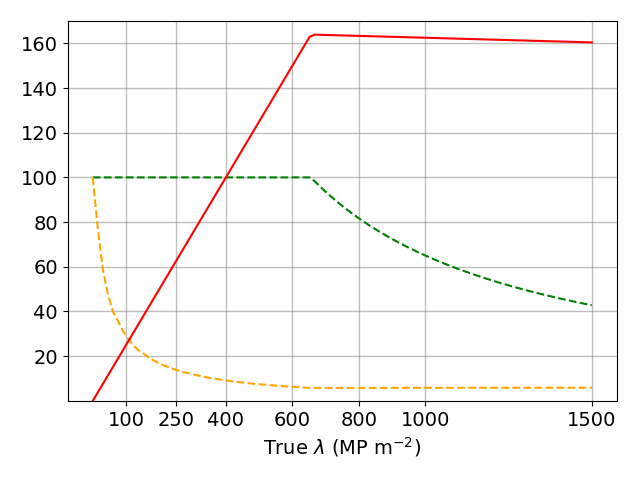}  \\
\end{tabular}
\end{center}
\caption{\small
\textbf{Upper panels:} Expected loss functions as a function of the number of quadrats $m$ (bottom axis, with $r_1 = 5 \cdot 10^{-5}$ and $r_2 = 3 \cdot 10^{-3}$, see text) and the corresponding sampled area $a = m A$ (top axis). The curves represent: the variance-reduction for abundance estimation $\Lambda$, $L_1^*( m | \alpha, \beta, A )$ (blue, dash–dot), the prior expected variance-reduction for polymer composition $\mathbf{P}$, $E[L_2^*( q \mid N , \mathbf{\gamma} )]$ (orange, dashed) and the composite design criterion, $L^*( m \mid A )$ in \eqref{eq:opti_L} (black, solid). Each panel includes an inset showing the prior for microplastic abundance $\Lambda$ used in that column. The asterisk marks the optimal design, $m^*$, i.e. the number of quadrants to be sampled, that minimizes $L^*$.
\textbf{Lower panels:} Performance of the optimal design under different true values of $\Lambda$; the expected variance-reduction for polymer composition, $100 E[L_2^*]$ (orange, dashed), the corresponding subsampling proportion $100 q$ (green, dashed), and the number of particles selected for polymer categorization, $\bar{n}(q) = \lfloor n q(n) \rfloor$ (red, solid).
\label{fig:Designs}}
\end{figure}

In Figure~\ref{fig:Designs}, we present two examples of the resulting cost functions in Equation~\eqref{eq:total_L} for two contrasting prior distributions for $\Lambda$, both with shape parameter $\alpha = 3$ but with mode at 200~MP~m$^{-2}$ (left panels, $\beta = 0.01$) and at 800~MP~m$^{-2}$ (right panels, $\beta = 0.0025$). The former represents a conservative prior in which low abundances are expected, although larger abundances are also foreseen up to 1,000~MP~m$^{-2}$ (see the inset in the top left panel of Figure~\ref{fig:Designs}). The latter represents a prior in which abundances up to 4,000~MP~m$^{-2}$ are foreseen (see the inset in the top right panel of Figure~\ref{fig:Designs}). That is, the expectation of a less polluted beach versus a possibly heavily polluted one. These results indicate that the optimal number of quadrants to be sampled is $m^* = 7$ and $m^* = 4$, respectively. That is, if we expect a more polluted beach, we may sample a smaller total area of the beach.

This is to be expected, as in a polluted beach, more MPs are expected in a smaller sampled area, aiming to reduce both the posterior variance for $\Lambda$, $L_1^*$, and the expected variance for $\mathbf{P}$, $E[L_2^*]$. Moreover, since less effort is put into sampling on the beach, more MPs may be categorized through polymer analysis, with $\bar{n}(q)$ reaching a maximum of nearly 180. This compares with a maximum of about 120 for $\bar{n}(q)$ in the first prior (less polluted beach), where $m^* = 7$ beach quadrants are sampled (see Figure~\ref{fig:Designs}, bottom panels). 

Note how, for low counts, the proportion $q$ of MPs to be categorised using polymer analysis is initially 1 (i.e., all suspected MPs are categorised), and around $n = 100$, $q$ starts to decrease, but $\bar{n}(q)$, the actual number of MPs to be categorized, remains approximately constant. This coincides with the current intuitive approach of the REMARCO method \citep{REMARCO2024}, although here it arises as a result of our formal analysis. This is a result of the constrained optimization in which it makes little sense to continue to categorize MPs.  This is seen in the U shape form of $E[L_2^*]$.  While $L_1^*$ logically decreases with the beach sampled are, $E[L_2^*]$ first decreases but, as more MPs are expected and less budget is available, $E[L_2^*]$ then increases, to return to 1, where all budget was spent in sampling in the beach and no MPs may be categorized.  The optimal point is a compromise between these two (see Figure~\ref{fig:Designs}, top panels).

\subsection{Robustness and sensitivity analysis}

In this section, we focus on the robustness of our experimental design to changes in prior assumptions, cost parameters, or available resources. In this section, we do not focus on the robustness of posterior inference for the parameters $\Lambda$ and $\mathbf{P}$ \textit{per se}, but rather examine how the optimal design itself adapts to variations in key inputs such as the total budget, the relative cost of categorization, and prior expectations of abundance. These analyses allow us to evaluate whether the proposed framework yields stable, interpretable, and practically useful outcomes across a range of realistic scenarios.

\subsubsection{Effect of Increasing the Cost of Polymer Analysis}

\begin{figure}[h!]
\begin{center}
\begin{tabular}{cc}
\includegraphics[scale=0.4]{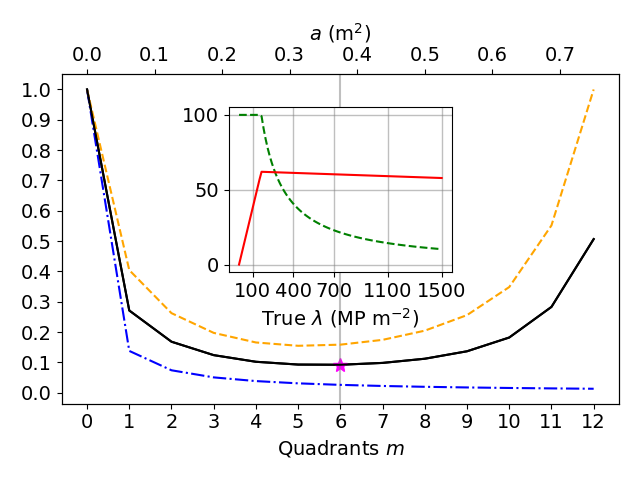} &
\includegraphics[scale=0.4]{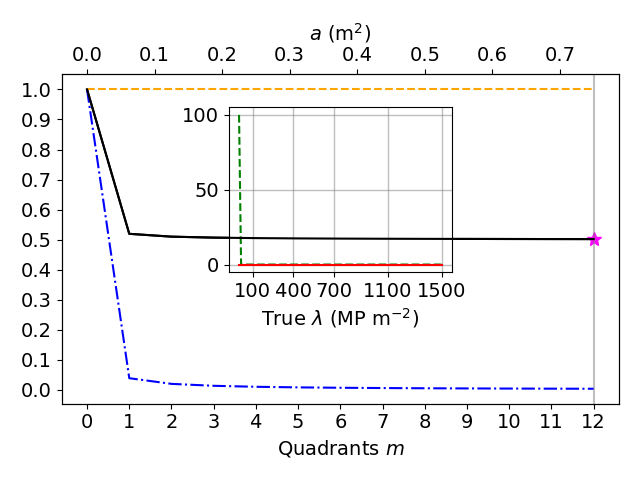} \\
\end{tabular}
\end{center}
\caption{
Effect of increasing the relative cost of polymer analysis on the optimal experimental design. \textbf{Left panel:} Optimal design under a scenario in which the cost of polymer categorization is doubled relative to the baseline $r_2 = 3 \cdot 10^{-3}$. As expected, the design shifts toward increased spatial sampling and a reduced number of particles selected for polymer analysis. This reflects the higher marginal cost of categorization and the greater relative gain in posterior precision for the abundance parameter $\Lambda$. \textbf{Right panel:} Extreme scenario in which the cost of polymer analysis is 1000 times higher than the baseline. The optimal design allocates all resources to maximizing the sampled area, choosing not to analyze any particles due to the prohibitive cost. The solution lies on the boundary of the feasible design space, suggesting that a different optimal balance may emerge if additional resources become available.
In this figure and Figures~\ref{fig:robust_budget} and~\ref{fig:robust_budget_inc} the inset is as in Figure~\ref{fig:Designs}'s lower panels.
}
\label{fig:robust_FTIR}
\end{figure}

We first consider how the optimal design responds when the cost of polymer categorization increases relative to the baseline value used in the previous section of $r_2 = 3 \cdot 10^{-3}$. Figure~\ref{fig:robust_FTIR} illustrates two scenarios: the left panel doubles the cost of polymer analysis, while the right panel considers a much higher cost, 1000 times the baseline.

As expected, the design adjusts accordingly. In the moderate case (left), resources shift away from categorization and toward increased spatial coverage. This reallocation reflects a higher marginal cost for categorization and the greater gain in posterior precision for the abundance parameter $\Lambda$. In the extreme case (right), the model recommends not performing polymer analysis at all, instead allocating the entire budget to spatial sampling and MPs counting. The solution lies on the boundary of the design space. Importantly, this outcome does not imply that categorization is uninformative in general; instead, it highlights the prohibitive cost under the given budget. We expect that with a larger budget, the optimal design would reintroduce categorization efforts, allocating some resources to polymer identification. Such behavior is both interpretable and desirable: the model avoids overcommitting resources when constraints are tight, but remains flexible when more resources become available.

\subsubsection{Effect of Budget Reductions}

We also examined how the optimal design adapts under more constrained conditions, specifically when the total available budget is reduced. Limited resources are a common feature in environmental monitoring campaigns, and the design framework must provide sensible recommendations even under tight constraints.

\begin{figure}[h!]
\begin{center}
\begin{tabular}{cc}
\includegraphics[scale=0.4]{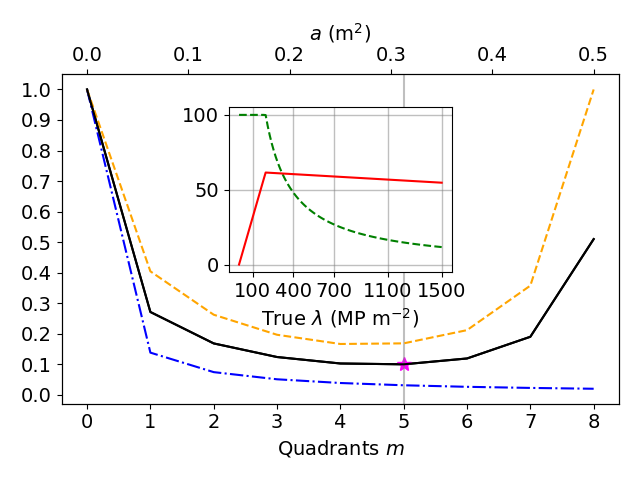} &
\includegraphics[scale=0.4]{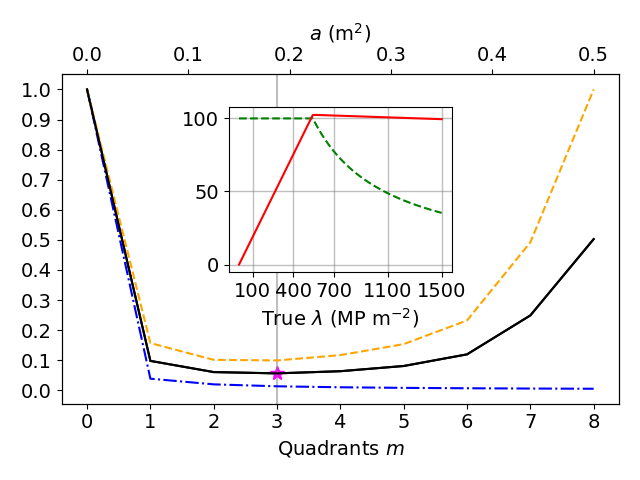} \\
\end{tabular}
\end{center}
\caption{
Effect of a reduced budget (corresponding to a maximum of 8 quadrats) on the optimal experimental design. Left panel: Optimal design under a prior with a lower expected abundance of MPs as in Figure~\ref{fig:Designs} top left panel. Right panel: Results under a prior with a much higher expected MPs abundance, as shown in the top right panel of  Figure~\ref{fig:Designs}.
}
\label{fig:robust_budget}
\end{figure}

Figure~\ref{fig:robust_budget} explores scenarios in which the total available budget is reduced, allowing a maximum of only eight quadrats to be sampled. The two panels show optimal designs under two different priors for $\Lambda$: a low expected abundance (left) and a high expected abundance (right).

In the low-abundance scenario, the design recommends sampling five quadrats and categorizing approximately 60 particles. Although this deviates from current REMARCO guidelines (which always assume a fixed number of 100 particles for polymer analysis~\citealt{REMARCO2024}), it represents a balanced compromise between spatial resolution and categorization depth under limited resources. In contrast, when the prior suggests a more polluted environment (right panel), the design is simplified to just three quadrats, reflecting the expectation that fewer samples will be sufficient to observe enough particles. The remaining budget is then allocated to categorize up to $\sim$100 particles, allowing meaningful inference about polymer composition. These shifts highlight how our design strategy tailors the design to both prior expectations and resource constraints.

\subsubsection{Effect of Budget Increases}

To complement the previous analyses under budget constraints, we now consider scenarios where a larger budget is available. Understanding how the optimal design scales with increasing resources is essential not only for planning realistic expansions in monitoring capacity but also for ensuring that additional investments translate into meaningful gains in information and inference quality.

\begin{figure}[h!]
\begin{center}
\begin{tabular}{cc}
\includegraphics[scale=0.4]{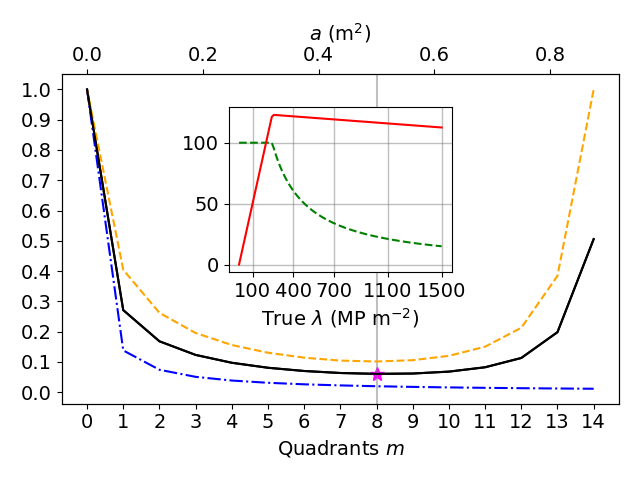} &
\includegraphics[scale=0.4]{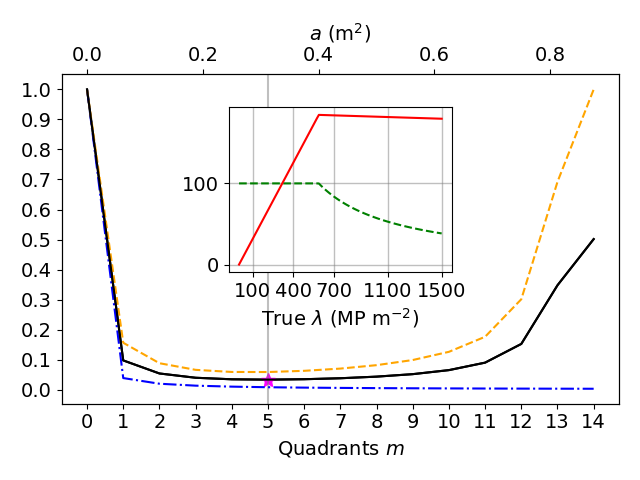} \\
\end{tabular}
\end{center}

\caption{
Effect of an increased budget (from a maximum of 12 to 14 quadrats) on the optimal experimental design.
Left panel: Optimal design under a prior with a lower expected abundance of MPs as in Figure~\ref{fig:Designs} top left panel. Right panel: Results under a prior with a much higher expected MPs abundance, as shown in the top right panel of Figure~\ref{fig:Designs}.
}
\label{fig:robust_budget_inc}
\end{figure}

Figure~\ref{fig:robust_budget_inc} illustrates how the design evolves as the available budget increases, from the equivalent of 12 to 14 quadrants. The left panel considers a moderate expected abundance, and the optimal design recommends sampling eight quadrat and categorizing around 125 particles. In the right panel, where the expected abundance is higher, the optimal design aligns with REMARCO's method \citep{REMARCO2024} in terms of spatial sampling. Still, the increased budget allows for additional investment in the categorization phase, recommending the analysis of nearly 200 suspected particles—almost doubling the categorization effort. In both cases, the designs indicate that increased resources should be allocated proportionally to both the sampling and analysis phases, depending on the expected MPs abundance.

These results demonstrate that, when additional resources are available, the design expands effort across both field sampling and laboratory analysis, adjusting the balance depending on prior cpmta,omatopm expectations. This proportional scaling reinforces the intuitive behavior of the proposed methodology: increased budgets should not automatically lead to uniform increases across all phases, but instead respond to the marginal utility of each design decision.

Overall, these scenarios demonstrate that the proposed design framework adapts predictably and sensibly across a wide range of practical constraints. By responding dynamically to both prior beliefs and resource limitations. Therefore, our design strategy provides a robust foundation for informed decision-making in MPs monitoring programs.

\section{Examples}\label{sec:examples}

\begin{figure}[h!]
\begin{center}
\begin{tabular}{cc}
\includegraphics[scale=0.5]{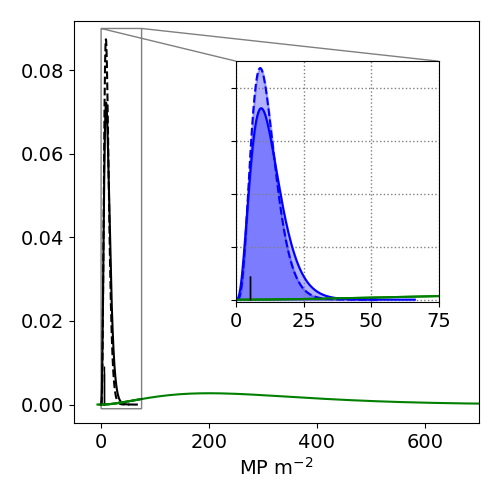} &
\includegraphics[scale=0.5]{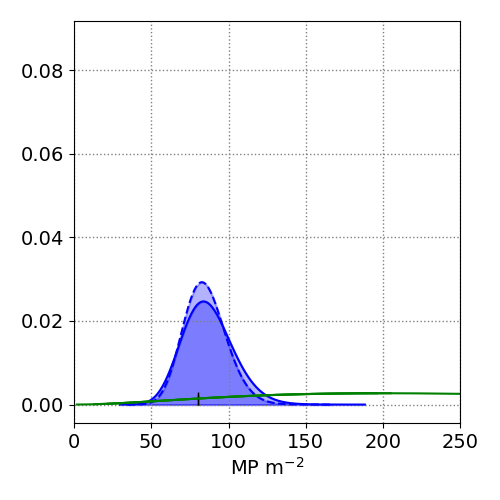} \\
\end{tabular}
\end{center}
\caption{\label{fig:LaPosts} 
Posterior of the abundance $\Lambda$ for ``true'' abundances 5 (left panel) and 80 (right panel), in MP~m$^{-2}$. These true values are flagged with a vertical line, and the same Gamma prior is used, with shape parameter $\alpha = 3$ and mode at 200~MP~m$^{-2}$ ($\beta = 0.01$), plotted in green. The standard design, with $m = 5$ quadrants (solid lines), is compared to our alternative design, with $m^* = 7$ quadrants (dashed lines). The expected number of observed MPs, $n$, is 1 and 2, respectively (left panel), and 25 and 35 (right panel).
}
\end{figure}

\begin{figure}[h!]
\begin{center}
\begin{tabular}{cc}
\includegraphics[scale=0.45]{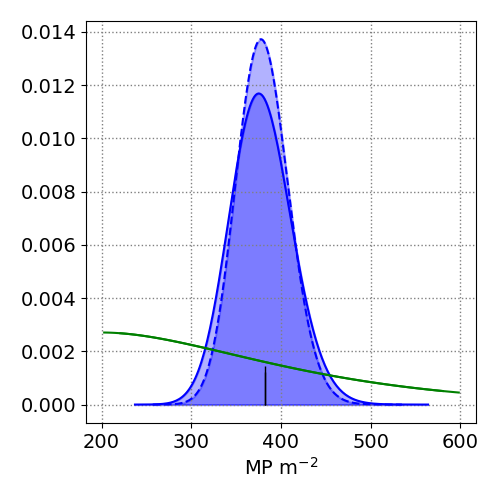} &
\includegraphics[scale=0.47]{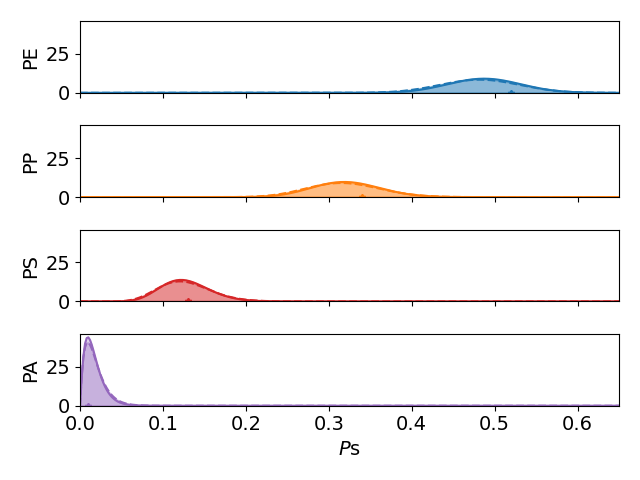} \\
\includegraphics[scale=0.45]{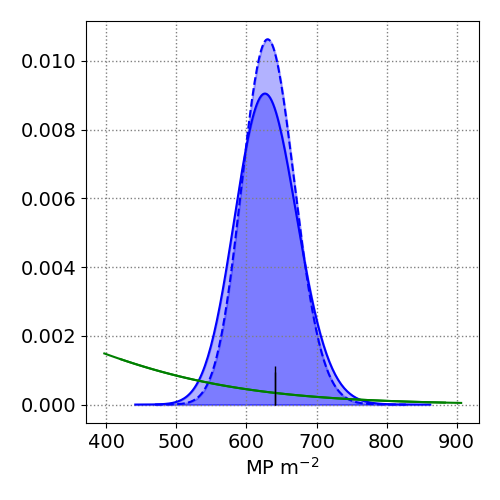} &
\includegraphics[scale=0.47]{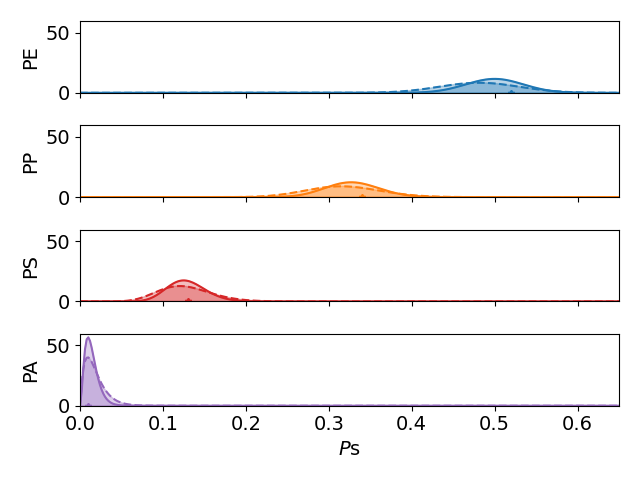} \\
\end{tabular}
\end{center}
\caption{\label{fig:Examples} 
Examples comparing $m = 5$ (solid lines) vs.\ $m^* = 7$ (dashed lines). Also, we take $\Lambda = 382$~MP~m$^{-2}$ (top panels) and $\Lambda = 600$~MP~m$^{-2}$ (bottom panels) as ``true'' values. In the former case, $n = 119$ vs.\ $n = 167$ are expected to be observed; however, in the standard design, all MPs are categorized, whereas in our alternative design, $q = 0.6$, as shown in Figure~\ref{fig:Designs} (bottom left panel). Nonetheless, the marginal posterior distributions of the $P_i$'s results are nearly identical, differing only in the categorization of 101 MPs. Similar, but more extreme, is the former case, where $n = 200$ vs.\ $n = 280$ but only 99 MPs ($q \approx 0.36$!) are categorized, nevertheless resulting in quite similar posteriors for the $P_i$'s.
}
\end{figure}

We present comparative examples of the standard design of sampling $m = 5$ quadrants from the REMARCO method \citep{REMARCO2024} versus the alternative design of sampling $m^* = 7$ quadrants on the beach, as shown in Figure~\ref{fig:Designs}, left panels. Results are presented in Figures~\ref{fig:LaPosts} and~\ref{fig:Examples}, respectively. In all cases, the posterior distributions are plotted with solid lines for the standard design ($m = 5$) and dashed lines for our alternative design ($m^* = 7$). We only use synthetic data, fixing a ``true'' value and using the expected observation as data, given the corresponding design.

The first example considers the posterior distribution of the abundance $\Lambda$, comparing the case of a very low abundance, $\Lambda = 5$~MP~m$^{-2}$, with a low abundance, $\Lambda = 80$~MP~m$^{-2}$. The second example considers a more contaminated beach and also includes MPs categorization with the corresponding posterior distribution for the polymer proportions $\mathbf{P}$.

We present the results of the first example (very low abundance) in Figure~\ref{fig:LaPosts}. Note that even with very low counts, there is substantial information, since an area of $a = 0.3125$~m$^2$ or $a = 0.4375$~m$^2$ is sampled (for $m = 5$ or $m^* = 7$, respectively). This results in a sharp decrease in variance from the prior, as seen in Figure~\ref{fig:LaPosts}. In fact, the expected reduction in variance for $\Lambda$ is $L_1^* = 0.031$ and $L_1^* = 0.022$, respectively; see Theorem~\ref{th:ExpVarLa}. Indeed, as we already know, since $\Lambda$ is Gamma-distributed, conclusions derived from these posteriors—such as Bayes estimators, Highest Density Intervals (HPIs), etc.—do not include negative values, in contrast with already mentioned current practices \citep[see][Figure 2, for example]{GarciaVarens2025}.

We present a second example based on a study by \citep[2025, Figure 6(b)][]{GarciaVarens2025}, for the ``T. Tomasa'' beach in south-central Cuba. Three sampling campaigns were conducted in different years using the standard sampling design of $m = 5$ quadrants (each with a standard sampling area of $A = 0.25^2$~m$^2$), with a total of 358 MPs found in $5 \cdot 3 = 15$ quadrants sampled. All MPs were categorized by FTIR, with resulting polymer type proportions of PE 52\%, PP 34\%, PS 13\%, and PA 1\%.

The original data are not available, and for this illustration, we assume a true $\Lambda = 358 / (15 \cdot 0.25^2) \approx 382$~MP~m$^{-2}$ and true polymer proportions equal to those observed, leaving the rest (see Table~\ref{tab:polymer_types}) with zero. Again, we compare the standard design of sampling $m = 5$ quadrants with the alternative design of sampling $m^* = 7$ quadrants, using the expected data as hypothetical observations in each design to compare the posteriors (see Figure~\ref{fig:Examples}, top panels).

A further illustration is presented with an even higher abundance of $\Lambda = 600$~MP~m$^{-2}$ (see Figure~\ref{fig:Examples}, bottom panels). The posterior distribution of $\Lambda$ (right panels) and the marginal posterior distributions of each $P_i$ for polymers with non-zero counts (left panels) are plotted in Figure~\ref{fig:Examples}.

In this last example, $q(a, n) < 1$ in both cases. In the first case, from $n = 167$ MPs observed, only 101 are categorized ($q(a, n) = 0.6$), and in the second case, from $n = 280$ MPs observed, only 99 are categorized ($q(a, n) \approx 0.36$). In this latter case, the effort of counting the suspected MPs becomes significant ($r_1 n$ in Equation~\eqref{eq:cost_0}), leaving less budget for categorization. In any case, this shows that categorizing more than 100 MPs starts to make little economic sense. From Theorem~\ref{th:ExpVarP}, we see that the expected reduction in variance for $\mathbf{P}$ is $L_2^*(q = 1 \mid n = 280, \gamma) = 0.0344$ versus $L_2^*(q = 0.36 \mid n = 280, \gamma) = 0.09$. Only a marginal gain would be obtained by categorizing all MPs.

The posterior for $\Lambda$ is more peaked for our design, given the higher area sampled; however, the posterior marginals for the $P_i$'s are identical or quite similar. This is a result of our optimization strategy, which produces a more rational use of resources, with results that we believe are both reasonable and plausible.

\section{Discussion and Conclusion}\label{sec:discussion}

This study presents a comprehensive Bayesian framework for the optimal experimental design of microplastic (MP) monitoring campaigns under realistic resource constraints. By formulating the design problem in terms of variance-based information gain, we offer an approach that prioritizes statistical efficiency. 

A key feature of our contribution is the two-stage design structure: first, determining the number of field samples to collect (quadrants in the context of beach sampling) and count suspect MPs, and second, allocating the remaining resources to categorize an appropriate proportion of the total observed MPs. This structure reflects the actual workflow in MP monitoring campaigns, allowing for coherent planning across both sampling and counting, and analysis phases.

Our Bayesian framework not only guides optimal design but also corrects for common statistical misuses. For example, previous studies have reported confidence intervals for MP abundance that include negative values—a clear indication of the misuse of inferential methods. By grounding our model in conjugate priors (Gamma-Poisson for abundance and Dirichlet-Multinomial for categorization), we provide a comprehensive probabilistic solution that yields internally coherent posterior distributions. As a result, our framework simultaneously defines a rational experimental design and ensures valid statistical inference as a natural consequence.

We further integrated a realistic cost model, capturing fixed expenditures, per-sample collection costs, and both particle counting and polymer categorization efforts. This model, normalized through dimensionless cost ratios, enables a clear comparison across sampling strategies, laboratories, and institutional contexts.

Through simulation-based experiments, we demonstrated the performance of the design under various prior assumptions and resource scenarios. Our results show that the optimal allocation dynamically adapts: in low-abundance settings or under tight budgets, the design priorities spatial coverage; in more polluted contexts, it reallocates resources to deepen polymer analysis without compromising overall inference quality. Moreover, the methodology can be easily adapted beyond beach environments and applied, with minor modifications, to other ecological monitoring problems, such as sampling marine sediments, peatlands, or ice cores.

Our examples and robustness analyses reveal that standard fixed designs (e.g., sampling five quadrats and categorizing all particles) are often suboptimal under budget constraints. In contrast, the proposed method provides quantitative and interpretable guidance tailored to specific objectives and limitations.

Looking ahead, an important direction for future research involves the role of homogeneity in MPs abundance at a site (i.e. may the samples from each quadrant be assumed to belong to the same distribution). In this work, we did not optimize for replication explicitly, but rather focused on total sampling effort. However, replicate samples are essential for assessing within-site variability, which in turn supports more robust inferences. This will be addressed in a forthcoming study focused on site homogeneity analysis, where the emphasis shifts toward evaluating spatial consistency across replicates. Such an analysis not only helps answer various scientific questions, such as the spatial stability of MPs, but also provides a stronger statistical foundation for developing and validating standardized monitoring guidelines. Also there is the possibility of including and intermediate step in the sequential analysis to decide on the maximum number of MPs for identification, before categorization. These considerations need to be left for future work.

In conclusion, this study builds upon the REMARCO monitoring guidelines \citep{REMARCO2024} by introducing a unified Bayesian framework that expands them into a resource-aware and statistically principled methodology. By formalizing both the design and analysis stages, we not only deliver an optimized experimental design for MPs monitoring, but also provide a coherent inferential framework for its interpretation.


\section*{Acknowledgments}
The authors would like to thank the PaleoStats group for providing the space to present and discuss the early ideas that led to this work. We also gratefully acknowledge the support of the 2023 Accelerate-C2D3 AI for Research and Innovation funding scheme at the University of Cambridge, which made the July 2024 PaleoStats meeting possible.

\section*{Funding}
MAL and JAC are partially founded by the grant ONRG N62909-24-1-2016. \\ 
MAL was funded by the 2023 Accelerate-C2D3 AI for Research and Innovation grant, which funded the 2024 PaleoStats meeting. \\ 
ACRF and JASC are partially funded by the grants from IAEA Technical Cooperation Regional Project RLA7028 and PAPIIT-DGAPA-UNAM IN109024.

\appendix

\section{Proof of Theorems \ref{th:ExpVarLa} and \ref{th:ExpVarP}}
\label{sec:appendix}

\subsection{Theorem \ref{th:ExpVarLa}, expected reduction of variance for $\Lambda$}

We resort to the prior predictive distribution of the $N_j$s to calculate the expectation of $L_1$ in \eqref{eq:loss_1}, which, using the Gamma posterior for $\Lambda$, is
\begin{equation*}
	L_1 ( m \mid \mathbf{N},\alpha,\beta, A) 
	= \frac{\beta^2}{\alpha} \text{Var}(\lambda | \alpha, \beta,\mathbf{N}) 
	=  \frac{\beta^2}{\alpha} \frac{\alpha + \sum_{j=1}^m N_j}{(\beta + mA)^2} .
\end{equation*}
In the usual way, the prior predictive distribution is obtained with $\int_0^\infty \text{Poi}(n_j \mid \lambda A) ~ \text{Gamma}(\lambda \mid \alpha, \beta) ~  d\lambda$ and is a Poisson-Gamma distribution \citep[p.119]{bernardo1994}.  In particular its expected value is
$E\left[ N_j \mid \alpha, \beta, A  \right] = A \frac{\alpha}{\beta}$.
Then
\begin{align}
	L_1^*( m | \alpha, \beta, A )
    &= \mathbb{E}_{\mathbf{N}} (L_1 ( m \mid \mathbf{N},\alpha,\beta, A) | \alpha, \beta) ) = \nonumber \\
    &=  \frac{\beta^2}{\alpha} \frac{ (\alpha + m \mathbb{E}[N_j\mid \alpha,\beta]) }{(\beta + mA)^2}  
    =  \frac{\beta^2}{\alpha} \frac{ (\alpha + m A \frac{\alpha}{\beta})}{(\beta + mA)^2} =
    \frac{1}{1 + \beta^{-1} m A}. \label{eq:lossfunction1}
\end{align}
Since the prior mean is $\lambda_0 = \frac{\alpha}{\beta}$ and the prior variance is $\nu_0 = \frac{\alpha}{\beta^2}$ then
\begin{align*}
	L_1^*( m | \alpha, \beta, A ) = \frac{1}{1+\frac{\nu_0}{\lambda_0} m A} 
\end{align*}
as desired.

\subsection{Theorem \ref{th:ExpVarP}, expected reduction of variance for $\mathbf{P}$}

We resort to the prior predictive distribution for $\mathbf{S}$ to calculate the expected loss of $L_2$ in \eqref{eq:L_star_p_nqc}. This prior predictive distribution is obtained with
$
    P(\mathbf{\bar{S}} = \mathbf{\bar{s}} \mid n,q , \boldsymbol{\gamma}) = \int f(\mathbf{s} \mid n, q , \mathbf{p}) \text{Dir}_k \left(\mathbf{p} \mid \boldsymbol{\gamma}, n, q \right) d\mathbf{p} . 
$
This integral simplifies to the Dirichlet-Multinomial distribution \citep{bernardo1994}.  In particular
\begin{equation}
\label{eqn:ExpVar_S}
\mathbb{E}\left[ S_i \mid n, q, \boldsymbol{\gamma} \right] = n \frac{\gamma_i}{\gamma_0} ~~\text{and}~~  \mathrm{Var}\left[ S_i \mid n, q, \boldsymbol{\gamma} \right] =
n \frac{\gamma_i}{\gamma_0}
\left( 1 - \frac{\gamma_i}{\gamma_0} \right)
\left( \frac{n + \gamma_0}{1 + \gamma_0} \right)
\end{equation}
where $\gamma_0 = \sum_{i=1}^k \gamma_i$.

Since 
$
L_2(q \mid \bar{\mathbf{s}}, n, \boldsymbol{\gamma}) 
= \frac{\text{Tr}(\Sigma_1)}{ \text{Tr}(\Sigma_0) }
$
letting $\theta_i = \gamma_i / \gamma_0$, and $\boldsymbol{\theta} = [\theta_1,\theta_2,\ldots,\theta_k]^T$ the covariance matrix of the Dirichlet distribution can then be expressed as
$
    \Sigma_0 = \frac{1}{1+\gamma_0} \left(  I_k \boldsymbol{\theta}  - \boldsymbol{\theta} \boldsymbol{\theta}^T \right) 
$
and taking the trace of this yields
$$
    \text{Tr}(\Sigma_0) = \frac{1}{1+\gamma_0} \left( \sum_{i=1}^k \theta_i - \sum_{i=1}^k \theta_i^2 \right)
    = \frac{1}{1+\gamma_0} \left( 1 - \sum_{i=1}^k \theta_i^2 \right) .
$$

This expression can be readily extended to the posterior distribution of $\mathbf{P}$,
namely, let
$\gamma_0^{(n)} = \gamma_0 + \bar{n}(q)$ and $\theta_i^{(n)} = \frac{\gamma_i + \bar{s}_i(q)}{\gamma_0 + \bar{n}(q)}$
then, the posterior covariance matrix is given by
\begin{align*}
    \Sigma_1 &= \frac{1}{1 + \gamma_0 + \bar{n}(q)} \left( I_k \boldsymbol{\theta}^{(n)} - \boldsymbol{\theta}^{(n)} {\boldsymbol{\theta}^{(n)}}^T \right) .
\end{align*}
Taking the trace, we obtain
\begin{align*}
    \text{Tr}(\Sigma_1) &= \frac{1}{1 + \gamma_0 + \bar{n}(q)} \left( 1 - \sum_{i=1}^k \left( \frac{\gamma_i + s_i}{\gamma_0 + \bar{n}(q)} \right)^2 \right)
\end{align*}
The variance reduction is therefore
\begin{align*}
	L_2(q \mid \mathbf{s}, n, \boldsymbol{\gamma}) &=
	\frac{
	    \frac{1}{1 + \gamma_0 + \bar{n}(q)} \left( 1 - \sum_{i=1}^k \left( \frac{\gamma_i + s_i}{\gamma_0 + \bar{n}(q)} \right)^2 \right)
	}{
	    \frac{1}{1 + \gamma_0} \left( 1 - \sum_{i=1}^k \left( \frac{\gamma_i}{\gamma_0} \right)^2 \right) 
	} .
\end{align*}
Let
$
D = 1 - \sum_{i=1}^k \left( \frac{\gamma_i}{\gamma_0} \right)^2
$
then
\begin{equation}
L_2(q \mid \mathbf{s}, n, \boldsymbol{\gamma}) =
\frac{1 + \gamma_0}{D\left(1 + \gamma_0 + \bar{n}(q)\right)} 
\left( 1 - \sum_{i=1}^k \left( \frac{\gamma_i + \bar{s}_i(q)}{\gamma_0 + \bar{n}(q)} \right)^2 \right).
\label{eq:L2star_simplified}
\end{equation}

From \eqref{eqn:ExpVar_S} it may be seen that
$$
\mathbb{E}\left[ \frac{\gamma_i + s_i}{\gamma_0 + \bar{n}(q)} \right] = \frac{\gamma_i}{\gamma_0} \label{eq:m_mean} ~~\text{and}~~
\mathrm{Var}\left[ \frac{\gamma_i + s_i}{\gamma_0 + \bar{n}(q)} \right] = \frac{ \bar{n}(q)(\gamma_0 - \gamma_i)\gamma_i }{ \gamma_0^2 (\gamma_0 + 1)(\gamma_0 + \bar{n}(q)) }
$$
and, applying the identity \( \mathbb{E}[X^2] = \mathrm{Var}(X) + (\mathbb{E}[X])^2 \), we have
\begin{align*}
\mathbb{E}\left[ \left( \frac{\gamma_i + \bar{s}_i(q)}{\gamma_0 + \bar{n}(q)} \right)^2 \right]
= \left( \frac{\gamma_i}{\gamma_0} \right)^2 + \frac{ \bar{n}(q)(\gamma_0 - \gamma_i)\gamma_i }{ \gamma_0^2 (\gamma_0 + 1)(\gamma_0 + \bar{n}(q)) }.
\end{align*}
Summing over we obtain
\begin{align*}
\sum_{i=1}^k \left( \frac{\gamma_i}{\gamma_0} \right)^2 &= 1 - D, \\
\sum_{i=1}^k \frac{ \bar{n}(q)(\gamma_0 - \gamma_i)\gamma_i }{ \gamma_0^2 (\gamma_0 + 1)(\gamma_0 + \bar{n}(q)) }
&= \frac{ \bar{n}(q) }{ \gamma_0^2 (\gamma_0 + 1)(\gamma_0 + \bar{n}(q)) } \left( \gamma_0^2 - \sum_{i=1}^k \gamma_i^2 \right) \\
&= \frac{ \bar{n}(q) D }{ (\gamma_0 + 1)(\gamma_0 + \bar{n}(q)) }.
\end{align*}
and hence, the expected value becomes
\begin{align}
\mathbb{E}_{\mathbf{S}}\left[ L_2(q \mid \mathbf{s}, n, \boldsymbol{\gamma}) \right]\nonumber
&= \frac{1 + \gamma_0}{D(1 + \gamma_0 + \bar{n}(q))} 
\left( 1 - (1 - D) - \frac{ \bar{n}(q) D }{ (\gamma_0 + 1)(\gamma_0 + \bar{n}(q)) } \right) \nonumber \nonumber\\
&= \frac{1 + \gamma_0}{1 + \gamma_0 + \bar{n}(q)} 
\left( 1 - \frac{ \bar{n}(q) }{ (\gamma_0 + 1)(\gamma_0 + \bar{n}(q)) } \right) \nonumber\\
&= \frac{(\gamma_0 + 1)(\gamma_0 + \bar{n}(q)) - \bar{n}(q)}{(1 + \gamma_0 + \bar{n}(q))(\gamma_0 + \bar{n}(q))}
= \frac{\gamma_0 + 1 - \frac{n}{\gamma_0 + n}}{\gamma_0 + 1 + n}.
\label{eq:expected_L2_corrected}
\end{align}
In the case of an uninformative prior with
$\gamma_i = \frac{1}{k}$ and $\gamma_0 = \sum_{i=1}^k \gamma_i = 1$ we get
\begin{align*}
L^*_2(q \mid n, \boldsymbol{\gamma}) =
\mathbb{E}_{\mathbf{S}}\left[ L_2(q \mid \mathbf{s}, n, \boldsymbol{\gamma}) \right]
&= \frac{2}{2 + \bar{n}(q)} \left( 1 - \frac{ \bar{n}(q) }{2(1 + \bar{n}(q))} \right) \\
&= \frac{2}{2 + \bar{n}(q)} \cdot \frac{2 + \bar{n}(q)}{2(1 + \bar{n}(q))} 
= \frac{1}{1 + \bar{n}(q)}
\end{align*}
which is the desired result.

\bibliographystyle{apalike}  
\bibliography{MP_biblio}  
\end{document}